\documentclass[journal, 12pt]{IEEEtran}
\usepackage{tcolorbox}
\tcbuselibrary{skins, breakable, theorems}

\usepackage{amssymb} 
\usepackage{pifont}  
\newcommand{\cmark}{\ding{51}} 
\newcommand{\xmark}{\ding{55}} 

\usepackage{multirow}
\usepackage{array}
\usepackage{graphicx}
\usepackage{float}
\usepackage{url}
\usepackage{xcolor}
\usepackage{amsmath}
\usepackage{hyperref}
\usepackage{cite}
\usepackage{amssymb}
\usepackage{booktabs}
\usepackage{colortbl}
\usepackage{caption}

\hypersetup{
    colorlinks = true,
    linkcolor = purple,
    citecolor = purple,
    anchorcolor = purple,
    filecolor = purple,
    urlcolor = purple
}

\colorlet{tableheadcolor}{gray!25} 
\colorlet{tablerowcolor}{gray!10} 
\title{A Contemporary Survey of  Large Language Model Assisted Program Analysis}
\author{Jiayimei Wang, Tao Ni, Wei-Bin Lee, Qingchuan Zhao$^{*}$%
\thanks{Jiayimei Wang, Tao Ni, and Qingchuan Zhao are with the Department of Computer Science, City University of Hong Kong, Hong Kong SAR (e-mail: jwang2664-c@my.cityu.edu.hk, taoni2@cityu.edu.hk, cs.qczhao@cityu.edu.hk). Wei-Bin Lee is with the Information Security Center, Hon Hai Research Institute, Taipei, Taiwan and the Department of Information Engineering and Computer Science, Feng Chia University, Taichung, Taiwan (e-mail: wei-bin.lee@foxconn.com, wblee@fcu.edu.tw).}
\thanks{$^{*}$ The corresponding author.}}

\begin{document}

\maketitle

\begin{abstract}
The increasing complexity of software systems has driven significant advancements in program analysis, as traditional methods unable to meet the demands of modern software development. To address these limitations, deep learning techniques, particularly Large Language Models (LLMs), have gained attention due to their context-aware capabilities in code comprehension. Recognizing the potential of LLMs, researchers have extensively explored their application in program analysis since their introduction. Despite existing surveys on LLM applications in cybersecurity, comprehensive reviews specifically addressing their role in program analysis remain scarce. In this survey, we systematically review the application of LLMs in program analysis, categorizing the existing work into static analysis, dynamic analysis, and hybrid approaches. Moreover, by examining and synthesizing recent studies, we identify future directions and challenges in the field. This survey aims to demonstrate the potential of LLMs in advancing program analysis practices and offer actionable insights for security researchers seeking to enhance detection frameworks or develop domain-specific models.
\end{abstract}




\section{Introduction}
With the continuous advancement of information technology, software plays an increasingly significant role in daily life, making its quality and reliability a critical concern for both academia and industry~\cite{software_challenges}.
This is because software vulnerabilities in domains such as finance, healthcare, critical infrastructure, aerospace, and cybersecurity~\cite{ref89} can lead to considerable financial losses or even societal harm~\cite{ref91}. Examples include data breaches in financial systems, malfunctioning medical devices, disruptions to power grids, failures in aviation control systems, and exploitation of security loopholes in sensitive government networks.
%
Accordingly, many techniques have been proposed to detect such vulnerabilities that compromise software quality and reliability, and program analysis has been proven effective in such tasks.
It aims to examine compter programs to identify or verify their properties to detect vulnerabilities through abstract interpretation, constraint solving, and automated reasoning~\cite{ref88}.
%

\begin{figure*}[t] 
    \centering 
    \includegraphics[width=\linewidth]{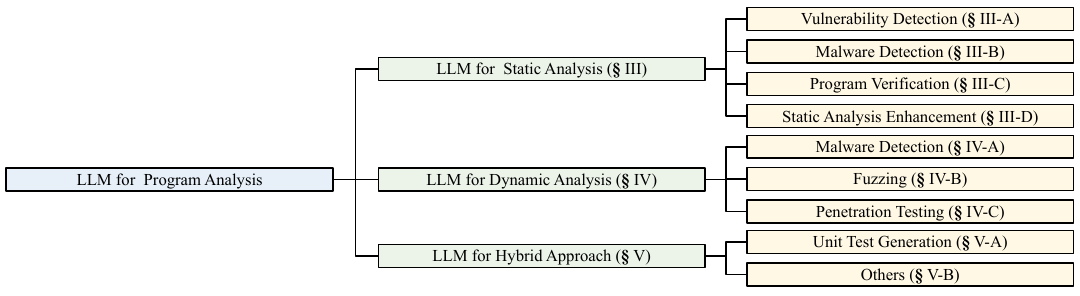}
    \vspace{-0.3in}
    \caption{Taxonomy of the survey.}
    \vspace{-0.3in}
    \label{fig:taxonomy} 
\end{figure*}

However, as software complexity and scale increase, traditional program analysis methods encounter challenges in meeting the demands of contemporary development. 
Specifically, these traditional methods face substantial challenges in handling dynamic behaviors, cross-language interactions, and large-scale codebases~\cite{ref93,ref94}. 
Fortunately, recent advancements in machine learning have initiated a shift in program analysis~\cite{ref96} and shed light on a promising research direction to address the limitations of traditional program analysis methods.

In particular, the literature has attempted to combine deep learning with program analysis, applying it to strengthen the detection of vulnerabilities and achieve automated code fixes, thereby minimizing human intervention and increasing precision~\cite{ref15}. 
However, deep learning models lack the ability to effectively integrate contextual information over long sequences, limiting their performance in tasks requiring deep reasoning or multi-turn understanding~\cite{ref17,ref18}. 
Consequently, these models struggle to handle complex software and large codebases and lack the capability for cross-project analysis. 
\looseness=-1

Fortunately, the most recent advancement, i.e., large language model (LLM), has been found promising in addressing the limitations of early deep learning models, such as constrained contextual understanding and generalization, enabling them to handle tasks across multiple domains with greater versatility~\cite{ref27, choi2025can, choi2025attributing, lin2023pushing, lin2024splitlora}.
Particularly, as for program analysis, LLMs surpass traditional deep learning methods and have been applied to various tasks~\cite{ref22, ni2024non, ni2023recovering, ni2023uncovering, fang2024automated}, including automated vulnerability and malware detection, code generation and repair, and providing scalable solutions that integrate static and dynamic analysis methods.
Moreover, it also shows a great potential to cope with the growing difficulty of analyzing modern software systems.


Though promising, the literature lacks a comprehensive and systematica view of LLM-assisted program analysis given the presence of numerous related attempts and applications.
Therefore, this work aims to systematically review the state-of-the-art of LLM-assisted program analysis applications and specify its role in the development of program analysis.
To this end, we systematically review the use of LLMs in program analysis and organized them into a structured taxonomy. \hyperref[fig:taxonomy]{Figure \ref{fig:taxonomy}} illustrates the classification framework, where the relevant research is categorized into LLM for static analysis, LLM for dynamic analysis, and hybrid approach. Unlike previous surveys that broadly examined the applications of LLMs in cybersecurity, our work narrows its focus to program analysis, delivering a more detailed and domain-specific exploration.
In addition, we collect the limitations mentioned in selected studies and analyze the improvements brought by the integration of LLMs, and specify the potential challenges and future research directions of LLMs in this domain.

The survey is organized as follows. We first introduce the background of program analysis and large language model in \autoref{sec:background}. We then examine the application of LLMs in static analysis in \autoref{sec:sa} and discusses the use of LLMs in dynamic analysis in \autoref{sec:da}.
We next explore how LLMs assist hybrid approaches that combine static and dynamic analysis in \autoref{sec:ha}.
We finally address the challenges of applying LLMs to program analysis and outline potential future research directions in \autoref{sec:discussion} and conclude the survey in \autoref{sec:conclusion}.

\section{Background}
\label{sec:background}

\looseness=-1
In this section, we first introduce prior knowledge about program analysis (\autoref{subsec:PA}), including static analysis and dynamic analysis and the limitations in existing approaches, and then present the concepts 
of LLMs as well as the necesseity of leveraging LLMs for advancing program analysis 
(\autoref{subsec:llm}).

\subsection{Program Analysis}
\label{subsec:PA}

\begin{figure*}[t] 
    \centering 
    \includegraphics[width=\linewidth]{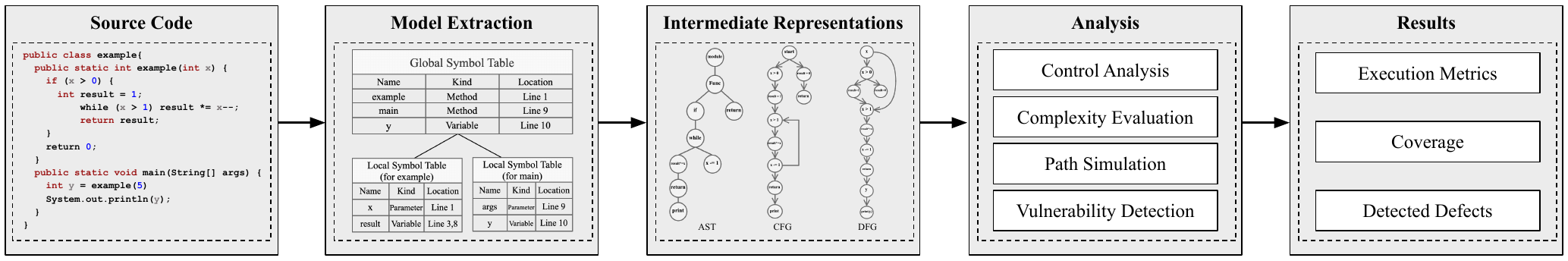} 
    \vspace{-0.2in}
    \caption{Static analysis workflow.}
    \vspace{-0.2in}
    \label{fig:SA} 
\end{figure*}

\begin{figure*}[t]
\centering
\includegraphics[width=0.9\textwidth]{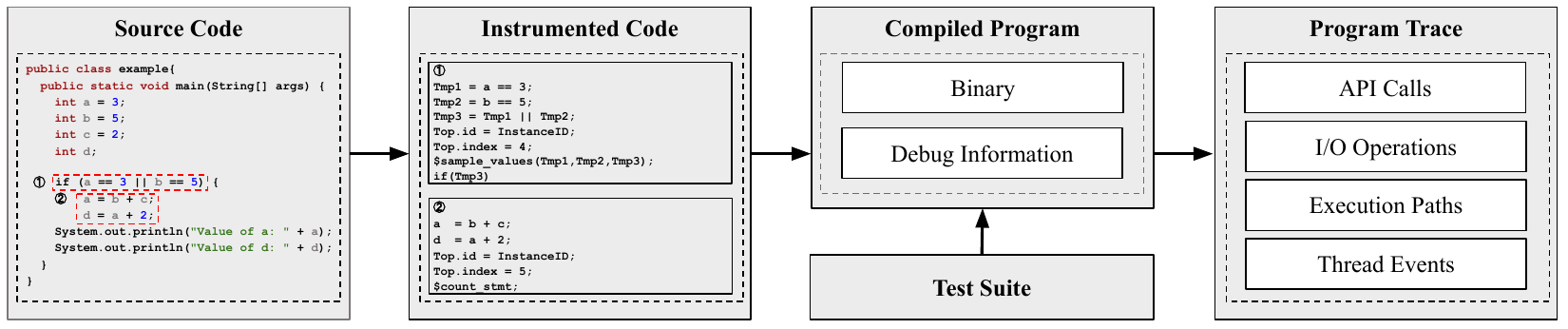}
\vspace{-0.1in}
\caption{Dynamic analysis workflow.}
\vspace{-0.3in}
\label{fig:DA}
\end{figure*}

Program analysis is the process of analyzing the behavior of computer programs to learn about their properties\cite{ref160}. Program analysis can find bugs or security vulnerabilities, such as null pointer dereferences or array index out-of-bounds errors. It is also used to generate software test cases, automate software patching and improve program execution speed through compiler optimization. 
Specifically, program analysis can be categorized into two main types: static analysis and dynamic analysis~\cite{ref3}. Static analysis examines a program's code without execution, dynamic analysis collects runtime information through execution, and hybrid analysis combines both approaches for comprehensive results.

\textbf{Static Analysis.} Static analysis (a.k.a. compile-time analysis) is a program analysis approach that identifies program properties by examining its source code without executing the program. The pipeline for static analysis consists of key stages illustrated in \autoref{fig:SA}. The process begins with parsing the code to extract essential structures and relationships, which are transformed into intermediate representations (IRs) such as symbol tables, abstract syntax trees (ASTs), control flow graphs (CFGs), and data flow graphs (DFGs). These IRs are then analyzed to detect issues such as unreachable code, data dependencies, and syntactic errors. These series of processes ultimately enhance code quality and reliability.

\textbf{Dynamic Analysis.} Dynamic analysis (a.k.a. run-time analysis) is a program analysis approach that uncovers program properties by repetitively executing programs in one or more runs~\cite{ref161}. The stages involved in dynamic analysis are depicted in \autoref{fig:DA}. These stages include instrumenting the source code to enable runtime tracking, compiling the instrumented code into a binary, and executing it with test suites. After completing the above steps, program traces such as function calls, memory accesses and system calls are captured.


\subsection{Large Language Models}
\label{subsec:llm}

Large Language Models (LLMs) are large-scale neural networks built on deep learning techniques, primarily utilizing the Transformer architecture~\cite{ref26}. Transformer models utilize self-attention mechanism to identify relationships between elements within a sequence, which enables them to outperform other machine learning models in understanding contextual relationships. Trained on vast datasets, LLMs learn syntax, semantics, context, and relationships within language, enabling them to generate and comprehend natural language~\cite{ref20}. Furthermore, LLMs possess knowledge reasoning capabilities, allowing them to retrieve and synthesize information from large datasets to answer questions involving common sense and factual knowledge. 

\looseness=-1
The architecture and configuration features of LLMs (e.g., model families, parameter size, and context window length) collectively determine their capabilities, performance and applicability. The studies selected in this survey involve LLM model families such as LLaMA~\cite{ref100}, CodeLLaMA~\cite{ref101} and GPT~\cite{ref102,ref103}. The parameter size of a large model typically refers to the number of variables used for learning and storing knowledge. The parameter size represents a model's learning capacity, indicating its ability to capture complexity and detail from data. Generally, larger parameter sizes enhance the model's expressive power, enabling it to learn more intricate patterns and finer details. The context window refers to the range of text fragments a model uses when generating each output. It determines the amount of contextual information the model can reference during generation. Selecting appropriate architectures and configurations for LLMs in different scenarios is crucial for optimizing their performance.


\begin{table*}[t]
\centering
\scriptsize
\setlength{\tabcolsep}{5pt}
\renewcommand{\arraystretch}{1.1}
\begin{tabular}{lcccccm{4.5cm}m{5cm}}
\toprule
\textbf{Reference} & \textbf{AST} & \textbf{CFG} & \textbf{DFG} & \textbf{OS} & \textbf{App} & \textbf{Vulnerability Type} & \textbf{LLM's assistance} \\ \midrule
LLift\cite{ref33} & \xmark & \cmark & \xmark & \cmark & \xmark & Use-before-initialization (UBI). & Path analysis. \\  \hline
SLFHunter\cite{ref39} & \cmark & \cmark & \cmark & \cmark & \xmark & Command injection vulnerabilities. & Taint sinks. \\  \hline
LATTE\cite{ref29} & \xmark & \cmark & \cmark & \cmark & \xmark & Binary taint analysis for data flows. & Binary taint analysis and code slicing. \\  \hline
IMMI\cite{ref61} & \xmark & \cmark & \cmark & \cmark & \xmark & Kernel memory bugs. & Memory allocation and deallocation intentions \\  \hline
DefectHunter\cite{ref28} & \cmark & \cmark & \cmark & \xmark & \cmark & General vulnerability. & Code sequence embeddings. \\  \hline
IRIS\cite{ref30} & \xmark & \cmark & \cmark & \xmark & \cmark & Taint analysis in smart contracts. & Taint sources and sinks. \\  \hline
VERACATION\cite{ref32} & \cmark & \xmark & \cmark & \xmark & \cmark & Syntactic-based vulnerability. & Filters non-vulnerability-related statements. \\  \hline
Mao \textit{et al.}\cite{ref34} & \cmark & \xmark & \cmark & \xmark & \cmark & Vulnerabilities in Code review processes. & Simulates multi-role discussions. \\  \hline
MSIVD\cite{ref31} & \xmark & \cmark & \cmark & \xmark & \cmark & General vulnerability. & Fine-tuned with multitask self-instructed learning. \\  \hline
GPTScan\cite{ref35} & \cmark & \xmark & \cmark & \xmark & \cmark & Smart contract logic vulnerabilities. & Analyzes smart contract semantics. \\  \hline
Yang \textit{et al.}\cite{ref36} & \cmark & \xmark & \xmark & \xmark & \cmark & IoT software vulnerability. & Explains vulnerabilities in code. \\  \hline
LLbezpeky\cite{ref37} & \xmark & \xmark & \xmark & \xmark & \cmark & Android security vulnerability. & Android application security. \\  \hline
SkipAnalyzer\cite{ref38} & \cmark & \xmark & \cmark & \xmark & \cmark & Bug detection. & Identifies bugs and generates patches. \\  \hline
HYPERION\cite{ref63} & \xmark & \cmark & \cmark & \xmark & \cmark & DApp Inconsistencies. & Extracts attributes of smart contract bytecode. \\  \hline
Zhang \textit{et al.}\cite{ref104} & \cmark & \xmark & \cmark & \xmark & \cmark & General vulnerability. & Detects vulnerabilities and fixes. \\  \hline
GPTLENS\cite{ref105} & \cmark & \xmark & \cmark & \xmark & \cmark & Smart contract vulnerability. & Generates diverse vulnerability hypotheses. \\  \hline
LuaTaint\cite{ref128} & \cmark & \cmark & \cmark & \xmark & \cmark & IoT vulnerability  & Prunes false alarms. \\
\bottomrule
\end{tabular}
\caption{Overview of the intermediate representations (AST, CFG, DFG) employed, their application domains (OS-level or application-level vulnerabilities), their application to specific vulnerability types, and the assistance provided by LLMs across selected studies}
\label{tab:vd}
\end{table*}

\section{LLM for Static Analysis}
\label{sec:sa}
Static analysis examines various objects, such as analyzing vulnerabilities and detecting malware in source code binary executables. 
Analyzing vulnerabilities in source code requires techniques like dependency analysis and taint tracking to trace the flow of sensitive data. On the other hand, Detecting malware focuses on control flow examination and behavior modeling to identify malicious patterns. Consequently, LLM assistance differs by program type and analysis purpose, which will be discussed in this section across four directions: (i) vulnerability detection (\autoref{subsec:llm_for_vd}), (ii) malware detection (\autoref{subsec:SAMD}), (iii) program verification (\autoref{subsec:llm_for_pv}), and (iv) static analysis enhancement (\autoref{subsec:llm_for_sae}).

\subsection{LLM for Vulnerability Detection}
\label{subsec:llm_for_vd}
Vulnerability detection focuses on identifying potential security risks or weaknesses in software through automated tools and techniques, which demand precise code analysis and a deep understanding of program behavior~\cite{chen2024attention, lu2023detecting}. Leveraging their advanced contextual comprehension, LLMs can analyze both semantic and syntactic patterns in source code, providing actionable suggestions and remediation strategies for addressing vulnerabilities. As a result, integrating LLMs into vulnerability detection has become a prominent application in program analysis.
\looseness=-1

\begin{figure*}[t] 
    \centering 
    \includegraphics[width=0.95\textwidth]{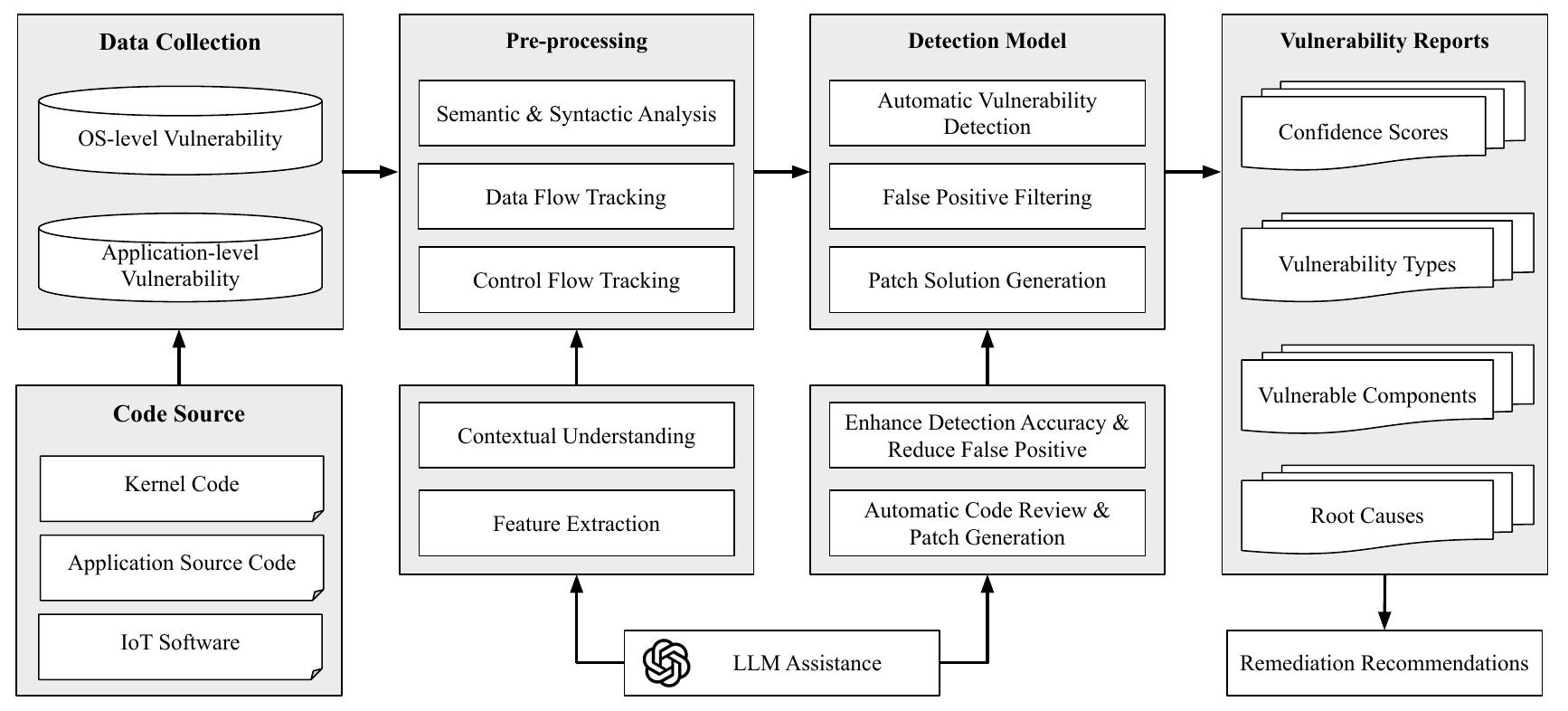}
    \vspace{-0.1in}
    \caption{A diagram of LLMs’ application in vulnerability detection. }
    \vspace{-0.25in}
    \label{fig:diagram1} 
\end{figure*}

To provide a clearer understanding of LLM applications in vulnerability detection, \autoref{tab:vd} summarizes the intermediate representations (IRs) utilized and the specific vulnerability types addressed in selected studies. \autoref{fig:diagram1} offers a visual overview of LLM integration at various stages, highlighting their roles in contextual understanding, feature extraction, enhanced detection accuracy, and remediation strategies. These capabilities enable efficient and precise identification of OS-level and application-level vulnerabilities. 
Additionally, a detailed comparison of the best-performing LLMs in the reviewed studies reveals key factors influencing their effectiveness and adoption. \autoref{tab:vd_llm} presents a comprehensive summary of these models, including their model family, parameter sizes, context window sizes, and open-source availability.
\looseness=-1

\textbf{OS-level Vulnerability. } OS-level vulnerabilities refer to security flaws within critical components of an operating system, such as the kernel, system libraries, or device drivers. These vulnerabilities can compromise the stability and security of the entire system, allowing attackers to gain unauthorized access, disrupt operations, or cause system-wide failures affecting all running applications. Common examples include memory management errors, privilege escalation, and resource misuse.
Leveraging LLMs, tools like the LLift framework~\cite{ref33} address challenges such as path sensitivity and scalability in detecting OS-level vulnerabilities. By combining constraint-guided path analysis with task decomposition, LLift improves the detection of issues like use-before-initialization (UBI) in large-scale codebases.
Ye~\textit{et al.}~\cite{ref39} developed SLFHunter, which integrates static taint analysis with LLMs to identify command injection vulnerabilities in Linux-based embedded firmware. The LLMs are utilized to analyze custom dynamically linked library functions and enhance the capabilities of traditional analysis tools. Furthermore, Liu~\textit{et al.}~\cite{ref29} proposed a system called LATTE, which combines LLMs with binary taint analysis. The code slicing and prompt construction modules serve as the core of LATTE, where dangerous data flows are isolated for analysis. These modules reduce the complexity for LLMs by providing context-specific input, allowing improved efficiency and precision in vulnerability detection through tailored prompt sequences that guide the LLM in the analysis process.
In addition, Liu~\textit{et al.}~\cite{ref61} proposed a system for detecting kernel memory bugs using a novel heuristic called Inconsistent Memory Management Intentions (IMMI). The system detects kernel memory bugs by summarizing memory operations and slicing code related to memory objects. It uses static analysis to infer inconsistencies in memory management responsibilities between caller and callee functions. LLMs assist in interpreting complex memory management mechanisms and enable the identification of bugs such as memory leaks and use-after-free errors with improved precision.
\looseness=-1

\textbf{Application-level Vulnerability. }Application-level vulnerabilities are security weaknesses found within individual software programs. These vulnerabilities can compromise the application's performance, data integrity, or user privacy. However, they typically do not affect the overall stability of the operating system. Common examples include input validation issues, logic errors, and misconfigurations. These vulnerabilities can result in unauthorized access or data breaches, as well as application-specific security incidents~\cite{zhao2019geo, zhao2020accuracy, ni2023eavesdropping, ni2023exploiting, ni2023xporter, ni2024sensor, zhao2020automatic}.

To address the challenges in application-level vulnerability detection, Wang~\textit{et al.}~\cite{ref28} introduced the Conformer mechanism, which integrates self-attention and convolutional networks to capture both local and global feature patterns. To further refine the detection process, they optimize the attention mechanism to reduce noise in multi-head attention and improve model stability. By combining structural information processing, pre-trained models, and the Conformer mechanism in a multi-layered framework, the approach improves detection accuracy and efficiency. Building on these advancements, IRIS~\cite{ref30} proposes a neuro-symbolic approach that combines LLMs with static analysis to support reasoning across entire projects. The static analysis is responsible for extracting candidate sources and sinks, while the LLM infers taint specifications for specific CWE categories. Similarly, Cheng~\textit{et al.}~\cite{ref32} combined semantic-level code clone detection with LLM-based vulnerability feature extraction. By integrating program slicing techniques with the LLM's semantic understanding, they refined vulnerability feature detection. This approach addresses the limitations of traditional syntactic-based analysis.

\begin{table}[ht]
\centering
\scriptsize
\setlength{\tabcolsep}{0.8pt}
\renewcommand{\arraystretch}{1.1}
\begin{tabular}{lllccc}
\toprule
\textbf{Reference} & \textbf{LLM} & \textbf{MF} & \textbf{Param} & \textbf{CW} & \textbf{Open-Source} \\ \midrule
LLift~\cite{ref33} & GPT-4-0613 & GPT-4 & - & 32768 & \xmark \\ \hline
SLFHunter~\cite{ref39} & GPT-4.0 & GPT-4 & - & 32768 & \xmark \\  \hline
LATTE~\cite{ref29} & GPT-4.0 & GPT-4 & - & 32768 & \xmark \\  \hline
IMMI~\cite{ref61} & ChatGPT-4-1106 & GPT-4 & - & 32768 & \xmark \\  \hline
DefectHunter~\cite{ref28} & UniXcoder & ~~~- & 250M & 768 & \cmark \\  \hline
IRIS~\cite{ref30} & GPT-4.0 & GPT-4 & - & 32768 & \xmark \\  \hline
VERACATION~\cite{ref32} & GPT-4.0 & GPT-4 & - & 1024 & \xmark \\  \hline
Mao~\textit{et al.}~\cite{ref34} & GPT-3.5-turbo & GPT-3.5 & 175B & 4096 & \xmark \\  \hline
MSIVD~\cite{ref31} & CodeLlama-13B & CodeLlama & 13B & 2048 & \cmark \\  \hline
GPTScan~\cite{ref35} & GPT-3.5-turbo & GPT-3.5 & 175B & 4096 & \xmark \\  \hline
Yang~\cite{ref36} & ChatGPT-4.0 & GPT-4 & - & 32768 & \xmark \\  \hline
LLbezpeky~\cite{ref37} & GPT-4.0 & GPT-4 & - & 32768 & \xmark \\  \hline
SkipAnalyzer~\cite{ref38} & ChatGPT-4.0 & GPT-4 & - & 8192 & \xmark \\  \hline
HYPERION~\cite{ref63} & LLaMA2\cite{ref108} & LLaMA & - & 4096 & \cmark \\  \hline
Zhang~\textit{et al.}~\cite{ref104} & ChatGPT-4.0 & GPT-4 & - & 8192 & \xmark \\  \hline
GPTLENS~\cite{ref105} & GPT-4.0 & GPT-4 & - & 32768 & \xmark \\  \hline
LuaTaint~\cite{ref128} & GPT-4.0 & GPT-4 & - & 1920 & \xmark \\
\bottomrule
\end{tabular}
\caption{Overview of the best-performing LLMs used in referenced papers, their model families (MF), parameter sizes (Param), context window sizes (CW), and open-source availability.}
\label{tab:vd_llm}
\end{table} 

Mao~\textit{et al.}~\cite{ref34} implemented a multi-role approach where LLMs act as different roles, such as testers and developers, simulating interactions in a real-life code review process. This strategy fosters discussions between these roles, enabling each LLM to provide distinct insights on potential vulnerabilities. MSIVD~\cite{ref31} introduces a multi-task self-instructed fine-tuning technique that combines vulnerability detection, explanation, and repair, improving the LLM's ability to understand and reason about code through multi-turn dialogues. Additionally, the system integrates LLMs with a data flow analysis-based GNN, which models the program's control flow graph to capture variable definitions and data propagation paths. This enables the model to rely not only on the literal information in the code but also on the program's graph structure for more precise detection. Similarly, GPTScan~\cite{ref35} demonstrates how GPT can be applied to code understanding and matching scenarios, reducing false positives and uncovering new vulnerabilities previously missed by human auditors.

In the domain of IoT software, Yang~\textit{et al.}~\cite{ref36} explored the application of LLMs combined with static code analysis for detecting vulnerabilities. By leveraging prompt engineering, LLMs enhance the efficiency of vulnerability detection and reduce costs, ultimately improving scalability and feasibility in large IoT systems~\cite{yuan2024itpatch, chen2022swipepass, zhao2022periscope}. Meanwhile, Xiang~\textit{et al.}~\cite{ref128} proposed LuaTaint, a static analysis framework designed to detect vulnerabilities in the web configuration interfaces of IoT devices. LuaTaint integrates flow-, context-, and field-sensitive static taint analysis with key features such as framework-specific adaptations for the LuCI web interface and pruning capabilities powered by GPT-4. By converting Lua code into ASTs and CFGs, the framework performs precise taint analysis to identify vulnerabilities like command injection and path traversal. The system uses dispatching rules and LLM-powered alarm pruning to improve detection precision, reduce false positives, and efficiently analyze firmware across large-scale datasets.

Mohajer~\textit{et al.}~\cite{ref38} presented SkipAnalyzer, a tool that employs LLMs for bug detection, false positive filtering, and patch generation. By improving the precision of existing bug detectors and automating patching, this approach significantly reduces false positives and ensures accurate bug repair. Meanwhile, Zhang~\textit{et al.}~\cite{ref104} introduced tailored prompt engineering techniques with GPT-4~\cite{ref103}, leveraging auxiliary information such as API call sequences and data flow graphs to provide structural and sequential context. This approach also employs chain-of-thought prompting to enhance reasoning capabilities, demonstrating improved accuracy in detecting vulnerabilities across Java and C/C++ datasets. Extending the application of LLMs in decentralized applications and smart contract analysis, Yang~\textit{et al.}~\cite{ref63} developed HYPERION, which combines LLM-based natural language analysis with symbolic execution to address inconsistencies between DApp descriptions and smart contracts. The system integrates a fine-tuned LLM to analyze front-end descriptions, while symbolic execution processes contract bytecode to recover program states, effectively identifying discrepancies that may undermine user trust.

\looseness=-1
For smart contract vulnerability detection, Hu~\textit{et al.}~\cite{ref105} introduced GPTLENS, a two-stage adversarial framework leveraging LLMs. GPTLENS assigns two synergistic roles to LLMs: an auditor generates a diverse set of vulnerabilities with associated reasoning, while a critic evaluates and ranks these vulnerabilities based on correctness, severity, and profitability. This open-ended prompting approach facilitates the identification of a broader range of vulnerabilities, including those that are uncategorized or previously unknown. Experimental results on real-world smart contracts show that GPTLENS outperforms traditional one-stage detection methods while maintaining low false positive rates. Focusing on Android security and software bug detection, Mathews~\textit{et al.}~\cite{ref37} introduced LLbezpeky, an AI-driven workflow that assists developers in detecting and rectifying vulnerabilities. Their approach analyzed Android applications, achieving over 90$\%$ success in identifying vulnerabilities in the Ghera benchmark. 

\begin{tcolorbox}[title = {Takeaway 1},
  fonttitle = \bfseries\normalsize, 
  fontupper = \rmfamily\normalsize, 
  fontlower = \itshape,
  breakable
  ]
  
Researchers utilize static analysis with different intermediate representations and LLMs to address different types of vulnerabilities. ASTs enhance syntactic reasoning and code representation for syntax-related vulnerabilities. CFGs address control flow issues such as privilege escalation by prioritizing paths and detecting anomalies. DFGs focus on data-flow vulnerabilities such as command injection, enabling LLMs to infer taint sources and refine detection rules. This integration of IRs and LLMs strengthens detection capabilities. Among LLMs, GPT-4 is commonly adopted for its large context window and versatility. Task-specific models like UniXcoder~\cite{unixcoder} perform well in specialized scenarios, while open-source models such as CodeLlama~\cite{codellama} provide reproducibility and flexibility.

\end{tcolorbox}

\subsection{LLM for Malware Detection}
\label{subsec:SAMD}
Malware detection determines whether a program has malicious intent and is an essential aspect of program analysis research. Initially, signature-based detection methods were predominantly used. As malware evolved, new detection techniques emerged, including behavior-based detection, heuristic detection, and model checking approaches. Data mining and machine learning algorithms soon followed, further enhancing detection capabilities~\cite{ref40, alasmary2018graph, shen2018android, alasmary2019analyzing, kang2015detecting, mohaisen2015amal}. 

\looseness=-1
Traditional malware detection methods struggle with challenges like obfuscation and polymorphic malware. LLMs offer a new approach to enhance detection accuracy and adapt to evolving threats by analyzing code semantics and patterns.
Fujii~\textit{et al.}~\cite{ref109} utilized decompiled and disassembled outputs of the Babuk ransomware as inputs to the LLM to generate function descriptions through carefully designed prompts. The generated descriptions were evaluated using BLEU~\cite{ref110} and ROUGE~\cite{ref111} metrics to measure functional coverage and agreement with analysis articles. Additionally, Simion~\textit{et al.}\cite{ref116} evaluated the feasibility of using out-of-the-box open-source LLMs for malware detection by analyzing API call sequences extracted from binary files. The study benchmarked four open-source LLMs (Llama2-13B, Mistral\cite{ref117}, Mixtral, and Mixtral-FP16~\cite{ref118}) using API call sequences extracted from 20,000 malware and benign files. The results showed that the models, without fine-tuning, achieved low accuracy and were unsuitable for real-time detection. These findings highlight the need for fine-tuning and integration with traditional security tools.

Analyzing malicious behaviors to detect malware is another approach. Zahan~\textit{et al.}~\cite{ref44} employed  a static analysis tool named CodeQL~\cite{codeql} to pre-screen npm packages. This step filtered out benign files, thereby reducing the number of packages requiring further investigation. Following this step, they utilized GPT-3 and GPT-4 models to analyze the remaining JavaScript code for detecting complex or subtle malicious behaviors. The outputs from the LLMs were refined iteratively. Accuracy improved through continuous adjustments to the model's focus based on feedback and re-evaluation.

Other studies focus on applying LLMs specifically to Android malware detection. Khan~\textit{et al.}~\cite{ref41} extracted Android APKs to obtain source code and opcode sequences, constructing call graphs to represent the structural relationships between functions. Models such as CodeBERT~\cite{CodeBERT} and GPT were employed to generate semantic feature representations, which were used to annotate the nodes in the call graphs. The graphs were enriched with structural and semantic information. These enriched graphs were then processed through a graph-based neural network to detect malware in Android applications. Zhao~\textit{et al.}~\cite{ref42} first extracted features from Android APK files using static analysis, categorizing them into permission view, API view, and URL $\&$ uses-feature view. A multi-view prompt engineering approach was applied to guide the LLM in generating textual descriptions and summaries for each feature category. The generated descriptions were transformed into vector representations, which served as inputs for a deep neural network (DNN)-based classifier to determine whether the APK was malicious or benign. Finally, the LLM produced a diagnostic report summarizing the potential risks and detection results.
\looseness=-1

\begin{tcolorbox}[title = {Takeaway 2},
  fonttitle = \bfseries\normalsize, 
  fontupper = \rmfamily\normalsize, 
  fontlower = \itshape,
  breakable]
The integration of LLMs with static analysis techniques enables the analysis of structured input sources, including decompiled functions, API call sequences, JavaScript code files, and APK attributes. A key commonality across approaches is the reliance on LLMs to process static features and generate semantic representations, textual descriptions, or embeddings, which are subsequently used for classification or detection tasks. Additionally, we notice that both open-source LLMs (e.g., Llama2-13B and Mistral) and proprietary models (e.g., GPT-4) are widely utilized in this task.
\end{tcolorbox}

\subsection{LLM for Program Verification}
\label{subsec:llm_for_pv}
Automated program verification employs tools and algorithms to ensure that a program's behavior aligns with predefined specifications, enhancing both software reliability and security. Traditional verification methods often require substantial manual effort, particularly for writing specifications and selecting strategies. These processes are often complex and prone to errors, especially in large-scale systems. In contrast, automated verification generates key elements such as invariants, preconditions, and postconditions, using techniques like static analysis and model checking to ensure correctness. The integration of LLMs further enhances this process by enabling the automatic analysis of code features and the efficient selection of verification strategies. This reduces manual intervention and significantly accelerates verification. Consequently, automated program verification has evolved into a more efficient and reliable method for ensuring software quality. This subsection introduces diverse applications of LLMs in program verification, highlighting their role in automating and enhancing critical tasks.

\autoref{tab:pv} provides an overview of various studies utilizing LLMs for program verification. It summarizes their targets, methodologies, and outcomes to highlight the diverse applications of these models in automating verification tasks. The inputs in these studies can be categorized into four types:
(i) Code, which includes program implementations or snippets used for analysis or synthesis.
(ii) Specifications, referring to formal descriptions of program behavior, such as preconditions, postconditions, or logical formulas.
(iii) Formal methods, encompassing mathematical constructs like theorems, proofs, and loop invariants for ensuring correctness.
(iv) Error and debugging information, such as counterexamples, type hints, or failed code generation cases that aid in resolving programming issues.
\looseness=-1

\begin{table*}[ht]
\centering
\scriptsize
\setlength{\tabcolsep}{5pt}
\renewcommand{\arraystretch}{1.1}
\begin{tabular}{l m{3cm} l c c m{2.5cm} m{4cm}}
\toprule
\textbf{Reference} & \textbf{Target} & \textbf{LLM} & \textbf{Param} & \textbf{OS} & \textbf{Input} & \textbf{Output} \\ \midrule
\multirow{4}{*}{\parbox[t]{2cm}{CoqPilot\cite{ref124}}} 
& \multirow{4}{*}{\parbox[t]{3cm}{Proof generation}} 
& Claude & - & \xmark 
& \multirow{4}{*}{\parbox[t]{3cm}{Formal methods}} 
& \multirow{4}{*}{\parbox[t]{4cm}{Coq proofs}} \\
& & LLaMA-2-13B & 13B & \cmark & & \\
& & GPT-3.5 & - & \xmark & & \\
& & GPT-4* & - & \xmark & & \\ \hline

\multirow{2}{*}{\parbox[t]{2cm}{Selene\cite{ref122}}} 
& \multirow{2}{*}{\parbox[t]{3cm}{Proof generation}} 
& GPT-3.5-turbo & 175B & \xmark 
& \multirow{2}{*}{\parbox[t]{3cm}{Specifications}} 
& \multirow{2}{*}{\parbox[t]{4cm}{Formal proofs}} \\
& & GPT-4* & - & \xmark & & \\ \hline

\multirow{2}{*}{\parbox[t]{2cm}{iRank\cite{ref54}}} 
& \multirow{2}{*}{\parbox[t]{3cm}{Loop invariant ranking}} 
& GPT-3.5-turbo & 175B & \xmark 
& \multirow{2}{*}{\parbox[t]{3cm}{Formal methods}} 
& \multirow{2}{*}{\parbox[t]{4cm}{Reranked LLM-generated invariants}} \\
& & GPT-4* & - & \xmark & & \\ \hline

Janßen \textit{et al.}\cite{ref125} 
& Loop invariant generation 
& GPT-3.5 & 175B & \xmark 
& Specifications 
& Valid  loop invariants \\ \hline

Pirzada \textit{et al.}\cite{ref126} 
& Loop invariant generation 
& GPT-3.5-Turbo-Instruct & 175B & \xmark 
& Formal methods
& Loop invariants \\ \hline

\multirow{3}{*}{\parbox[t]{2cm}{LaM4Inv\cite{ref127}}} 
& \multirow{3}{*}{\parbox[t]{3cm}{Loop invariant generation}} 
& LLaMA-3-8B & 8B & \cmark 
& \multirow{3}{*}{\parbox[t]{3cm}{Code}} 
& \multirow{3}{*}{\parbox[t]{4cm}{Loop invariants}} \\
& & GPT-3.5-Turbo & 175B & \xmark & & \\
& & GPT-4-Turbo* & - & \xmark & & \\ \hline

Pei \textit{et al.}\cite{ref79} 
& Invariant prediction 
& GPT-4 & - & \xmark 
& Code 
& Static invariants \\ \hline

\multirow{2}{*}{\parbox[t]{2cm}{AutoSpec\cite{ref53}}} 
& \multirow{2}{*}{\parbox[t]{3cm}{Specification synthesis}} 
& GPT-3.5-turbo-0613* & 175B & \xmark 
& \multirow{2}{*}{\parbox[t]{3cm}{Code}} 
& \multirow{2}{*}{\parbox[t]{4cm}{Specifications}} \\
& & Llama-2-70B & 70B & \cmark & & \\ \hline

\multirow{2}{*}{\parbox[t]{2cm}{LEMUR\cite{ref55}}} 
& \multirow{2}{*}{\parbox[t]{3cm}{Automated verification}} 
& GPT-3.5-turbo & 175B & \xmark 
& \multirow{2}{*}{\parbox[t]{3cm}{Specifications}} 
& \multirow{2}{*}{\parbox[t]{4cm}{Loop invariants}} \\
& & GPT-4* & - & \xmark & & \\ \hline

SynVer\cite{ref121} 
& Automated verification 
& GPT-4 & - & \xmark 
& Specifications 
& Candidate C programs \\ \hline

PropertyGPT\cite{ref123} 
& Smart contract verification 
& GPT-4-0125-preview & - & \xmark 
& Code and specifications 
& Formal verification properties \\ \hline

\multirow{2}{*}{\parbox[t]{2cm}{LLM-Sym\cite{ref67}}} 
& \multirow{2}{*}{\parbox[t]{3cm}{Python symbolic execution}} 
& GPT-4o-mini & - & \xmark 
& Error and debugging 
& Initial Z3Py code \\ 
& & GPT-4o & - & \xmark 
&  Error and debugging  
& Refined Z3Py code  \\ \hline

CFStra\cite{ref51} 
& Verification strategy selection
& GPT-3.5-turbo & 175B & \xmark 
&  Code and specifications
& Identified code features \\ \hline

Chapman \textit{et al.}\cite{ref64}
& Error specification inference 
& GPT-4 & - & \xmark 
&  Formal methods 
& Error specifications \\ \bottomrule
\end{tabular}
\caption{Overview referenced studies, detailing their targets, LLMs employed, parameter sizes (Param), open-source availability (OS), input types, and resulting outputs.}
\label{tab:pv}
\end{table*}

\textbf{Proof Generation.} Proof generation in program verification automates the creation of formal proofs to ensure program correctness, logical consistency, and compliance with specifications. This process reduces the need for manual effort and enhances verification efficiency by streamlining complex proof tasks. Kozyrev~\textit{et al.}~\cite{ref124} developed CoqPilot, a VSCode plugin that integrates LLMs such as GPT-4, GPT-3.5, LLaMA-2~\cite{ref100}, and Anthropic Claude~\cite{claude} with Coq-specific tools like CoqHammer~\cite{coqhammer} and Tactician~\cite{tac} to automate proof generation in the Coq theorem prover. The authors implemented premise selection for better LLM prompting and created an LLM-guided mechanism that attempted fixing failing proofs with the help of the Coq’s error messages. 
Additionally, Zhang~\textit{et al.}~\cite{ref122} developed the Selene framework to automate proof generation in software verification using LLMs. The framework is built on the industrial-level operating system microkernel~\cite{mic}, seL4~\cite{seL4Website}, and introduces the technique of lemma isolation to reduce verification time. Its key contributions include efficient proof validation, dependency augmentation, and showcasing the potential of LLMs in automating complex verification tasks. 

\textbf{Invariant Generation.} Invariant generation identifies properties that remain true during program execution, providing a logical foundation for verifying correctness and analyzing complex iterative structures like loops and recursion. 

Some studies have explored various ways to leverage LLMs for generating and ranking loop invariants. Janßen~\textit{et al.}~\cite{ref125} investigated the utility of ChatGPT in generating loop invariants. The authors used ChatGPT to annotate 106 C programs from the SV-COMP Loops category~\cite{SV-COMP} with loop invariants written in ACSL~\cite{acsl}, evaluating the validity and usefulness of these invariants. They integrated ChatGPT with the Frama-C~\cite{Framac} interactive verifier and the CPAchecker~\cite{ref52} automatic verifier to assess how well the generated invariants enable these tools to solve verification tasks. Results showed that ChatGPT can produce valid and useful invariants for many cases, facilitating software verification by augmenting traditional methods with insights provided by LLMs.
Additionally, Chakrabor~\textit{et al.}~\cite{ref54} observed that employing LLMs in a zero-shot setting to generate loop invariants often led to numerous attempts before producing correct invariants, resulting in a high number of calls to the program verifier. To mitigate this issue, they introduced iRank, a re-ranking mechanism based on contrastive learning, which effectively distinguishes correct from incorrect invariants. This method significantly reduces the verification calls required, improving efficiency in invariant generation.
\looseness=-1

Besides, Pei~\textit{et al.}~\cite{ref79} explored using LLMs to predict program invariants that were traditionally generated through dynamic analysis. By fine-tuning LLMs on a dataset of Java programs annotated with invariants from the Daikon~\cite{ref80} dynamic analyzer, they developed a static analysis-based method using a scratchpad approach. This technique incrementally generates invariants and achieves performance comparable to Daikon without requiring code execution. It also provides a static and cost-effective alternative to dynamic analysis.

\looseness=-1
Integrating LLMs with Bounded Model Checking (BMC) has shown potential in enhancing loop invariant generation. 
Pirzada~\textit{et al.}~\cite{ref126} proposed a modification to the classical BMC procedure that avoids the computationally expensive process of loop unrolling by transforming the CFG. Instead of unrolling loops, the framework replaces loop segments in the CFG with nodes that assert the invariants of the loop. These invariants are generated using LLMs and validated for correctness using a first-order theorem prover. This transformation produces loop-free program variants in a sound manner, enabling efficient verification of programs with unbounded loops. Their experimental results demonstrate that the resulting tool, ESBMC ibmc, significantly improves the capability of the industrial-strength software verifier ESBMC~\cite{esmbc}, verifying more programs compared to state-of-the-art tools such as SeaHorn~\cite{seahorn} and VeriAbs~\cite{veriabs}, including cases these tools could not handle.  
Wu~\textit{et al.}~\cite{ref127} proposed LaM4Inv, a framework that integrates LLMs with BMC to improve this process. The framework employs a 'query-filter-reassemble' pipeline. LLMs generate candidate invariants, BMC filters out incorrect predicates, and valid predicates are iteratively refined and reassembled into invariants.

\textbf{Automated Program Verification.} Automating program specification presents challenges such as handling programs with complex data types and code structures. To address these issues, Wen~\textit{et al.}~\cite{ref53} introduced an approach called AutoSpec. Driven by static analysis and program verification, AutoSpec uses LLMs to generate candidate specifications. Programs are decomposed into smaller components to help LLMs focus on specific sections. The generated specifications are iteratively validated to minimize error accumulation. This process enables AutoSpec to handle complex code structures, such as nested loops and pointers, making it more versatile than traditional specification synthesis techniques. 
Wu~\textit{et al.}~\cite{ref55} introduced the LEMUR framework. In this hybrid system, LLMs generate program properties like invariants as sub-goals, which are then verified and refined by reasoners such as CBMC~\cite{ref56}, ESBMC~\cite{esmbc} or UAUTOMIZER~\cite{ref58}. The framework is based on a sound proof system, thus ensuring correctness when LLMs propose incorrect properties. An oracle-based refinement mechanism improves these properties, enabling LEMUR to enhance efficiency in verification and handle complex programs more effectively than traditional tools.
Additionally, Mukherjee~\textit{et al.}~\cite{ref121} introduced SynVer, a framework that integrates LLMs with formal verification tools for automating the synthesis and verification of C programs. SynVer takes specifications in Separation Logic, function signatures, and input-output examples as input. It leverages LLMs to generate candidate programs and uses SepAuto, a verification backend, to validate these programs against the specifications. The framework prioritizes recursive program generation, reducing the dependency on manual loop invariants and improving verification success rates. 

\textbf{Others.} Other applications of LLMs in program verification include smart contract verification, symbolic execution, strategy selection and error specification inference. 
For instance, Liu~\textit{et al.}~\cite{ref123} developed a novel framework named PropertyGPT, leveraging GPT-4 to automate the generation of formal properties such as invariants, pre-/post-conditions, and rules for smart contract verification. The framework embeds human-written properties into a vector database and retrieves reference properties for customized property generation, ensuring their compilation, appropriateness, and runtime verifiability through iterative feedback and ranking. Similarly, Wang~\textit{et al.}~\cite{ref67} introduced an iterative framework named LLM-Sym. This tool leverages LLMs to bridge the gap between program constraints and SMT solvers. The process begins by extracting control flow paths, performing type inference, and iteratively generating Z3~\cite{z3} code to solve path constraints. A notable feature of LLM-Sym is its self-refinement mechanism, which utilizes error messages to debug and enhance the generated Z3 code. If the code generation process fails, the system directly employs LLMs to solve the constraints. Once constraints are resolved, Python test cases are automatically generated from Z3's outputs.
\looseness=-1

Another approach~\cite{ref51} automates the selection of verification strategies to overcome limitations of traditional tools like CPAchecker~\cite{ref52}. These tools often require users to manually select strategies, making the process more complex and time-consuming. LLMs analyze code features to identify suitable strategies, streamlining the verification process and minimizing user input. This automation not only improves efficiency but also minimizes reliance on expert knowledge.
Additionally, Chapman~\textit{et al.}~\cite{ref64} proposed a method that combines static analysis with LLM prompting to infer error specifications in C programs. Their system queries the LLM when static analysis encounters incomplete information, enhancing the accuracy of error specification inference. This approch is effective for third-party functions and complex error-handling paths.

\begin{tcolorbox}[title = {Takeaway 3},
  fonttitle = \bfseries\normalsize, 
  fontupper = \rmfamily\normalsize, 
  fontlower = \itshape,
  breakable]
The applications of LLMs in program verification span various tasks, including proof generation, specification synthesis, loop invariant generation, and strategy selection. These methods streamline the verification process by automating the generation of properties, invariants, and other critical components essential for program analysis. Despite their diverse applications, these methods share a common goal: reducing reliance on expert knowledge and improving verification efficiency. A key aspect of achieving this goal is the iterative refinement of LLM-generated outputs. This refinement process often incorporates static analysis or hybrid frameworks that integrate formal verification tools, further enhancing reliability.
\end{tcolorbox}

\subsection{LLM for Static Analysis Enhancement}
\label{subsec:llm_for_sae}
Beyond the previously mentioned applications of LLMs, other studies focus on leveraging LLMs to assist in certain processes of static analysis.

\textbf{Code Review Automation.} Lu~\textit{et al.}~\cite{ref65} proposed LLaMA-Reviewer, a model that leverages LLMs to automate code review. It incorporates instruction-tuning of a pre-trained model and employs Parameter-Efficient Fine-Tuning techniques to minimize resource requirements. The system automates essential code review tasks, including predicting review necessity, generating comments, and refining code.

\textbf{Code Coverage Prediction.} Dhulipala~\textit{et al.}~\cite{ref66} introduced CodePilot, a system that integrates planning strategies and LLMs to predict code coverage by analyzing program control flow. CodePilot first generates a plan by analyzing program semantics, dividing the code into steps derived from control flow structures, such as loops and branches. Subsequently, CodePilot adopts either a single-prompt approach (Plan+Predict in one step) or a two-prompt approach (planning first, followed by coverage prediction). These approaches guide LLMs to predict which parts of the code are likely to be executed based on the formulated plan.

\textbf{Decompiler Optimization.} Hu~\textit{et al.}~\cite{ref60} proposed DeGPT, a framework designed to enhance the clarity and usability of decompiler outputs for reverse engineering tasks. DeGPT begins by analyzing the raw output of decompilers, identifying issues such as ambiguous variable names, missing comments, and poorly structured code. The framework leverages LLMs in three distinct roles:Referee, Advisor, and Operator to propose and implement optimizations while preserving semantic correctness.

\textbf{Explainable Fault Localization.} Yan~\textit{et al.}~\cite{ref131} proposed CrashTracker, a hybrid framework that combines static analysis with LLMs. This approach improves the accuracy and explainability of crashing fault localization in framework-based applications. CrashTracker introduces Exception-Thrown Summaries (ETS) to represent fault-inducing elements in the framework. It also uses Candidate Information Summaries (CIS) to extract relevant contextual information for identifying buggy methods. ETS models are employed to identify potential buggy methods. LLMs then generate natural language fault reports based on CIS data, enhancing the clarity of fault explanations. CrashTracker demonstrates state-of-the-art performance in precision and explainability when applied to Android applications.

\textbf{Extract Method Refactoring.} Pomian~\textit{et al.}~\cite{ref129} introduced EM-Assist, a tool that combines LLMs and static analysis to enhance Extract Method (EM) refactoring in Java and Kotlin projects. EM-Assist uses LLMs to generate EM refactoring suggestions and applies static analysis to discard irrelevant or impractical options. To improve the quality of suggestions, the tool employs program slicing and ranking mechanisms to prioritize refactorings aligned with developer preferences. EM-Assist automates the entire refactoring process by leveraging the IntelliJ IDEA platform to safely implement changes.

\textbf{Obfuscated Code Disassembly.} Rong~\textit{et al.}~\cite{ref62} introduced DISASLLM, a framework that combines traditional disassembly techniques with LLMs. The LLM component validates disassembly results and repairs errors in obfuscated binaries, enhancing the quality of the output. Through batch processing and GPU parallelization, DISASLLM achieves substantial improvements in both the accuracy and speed of decoding obfuscated code, outperforming state-of-the-art methods

\textbf{Privilege Variable Detection.} Wang~\textit{et al.}~\cite{ref59} presented a hybrid workflow that combines LLMs with static analysis to detect user privilege-related variables in programs. The program is first analyzed to identify relevant variables and their data flows, which provides an initial set of potential user privilege-related variables. The LLM is used to evaluate these variables by understanding their context and scoring them based on their relationship to user privileges. 

\looseness=-1
\textbf{Static Bug Warning Inspection.} Wen~\textit{et al.}~\cite{ref132} proposed LLM4SA, a framework that integrates LLMs with static analysis tools to automatically inspect large volumes of static bug warnings. LLM4SA first extracts bug-relevant code snippets using program dependence traversal. It then formulates customized prompts with techniques such as Chain-of-Thought reasoning and few-shot learning. To ensure precision, the framework applies pre- and post-processing steps to validate the results. This approach tackles challenges like token limitations by optimizing input size, reduces inconsistencies in LLM responses through structured prompt engineering, and mitigates false positives via comprehensive validation. 

\textbf{Static Analysis Alert Adjudication.} Flynn~\textit{et al.}~\cite{ref68} proposed using LLMs to automatically adjudicate static analysis alerts. The system generates prompts with relevant code and alert details, enabling the LLM to classify alerts as true or false positives. To address context window limitations, the system summarizes relevant code and provides mechanisms for the LLM to request additional details or verify its classifications.

\textbf{Static Analysis Enhancement by Pseudo-code Execution.} Hao~\textit{et al.}~\cite{ref69} presented E$\&$V, a system designed to enhance static analysis using LLMs by simulating the execution of pseudo-code and verifying the results without needing external validation. It validates the results of the analysis through an automatic verification process that checks for errors and inconsistencies in the pseudo-code execution. This system is particularly useful for tasks like crash triaging and backward taint analysis in large codebases like the Linux kernel.

\begin{tcolorbox}[title = {Takeaway 4},
  fonttitle = \bfseries\normalsize, 
  fontupper = \rmfamily\normalsize, 
  fontlower = \itshape,
  breakable]
The methods in this subsection demonstrate how LLMs integrate with static analysis across domains such as debugging, fault localization, code refactoring, and privilege detection. A notable insight is the use of LLMs not just as generative tools but as collaborators that complement static analysis through contextual reasoning and iterative refinement. 
\end{tcolorbox}

\section{LLM for Dynamic Analysis}
\label{sec:da}
Dynamic analysis encompasses profiling and testing. Profiling focuses on understanding program performance by analyzing execution, such as counting statement or procedure executions through instrumentation. Testing aims to make sure the test suites can cover a program. Statement coverage verifies that every statement in the code is executed at least once during testing. Branch, condition, and path coverage evaluate how thoroughly all branches, conditions, and execution paths are tested~\cite{ref161}. This section examines how LLMs enhance dynamic analysis, focusing on (i) malware detection (\autoref{subsec:llm_for_dmd}) under profiling, (ii) fuzzing (\autoref{subsec:fuzzing}) and (iii) penetration testing (\autoref{subsec:llm_for_pt}) under testing.

\begin{table*}[]
\centering
\scriptsize
\setlength{\tabcolsep}{5pt}
\renewcommand{\arraystretch}{1.0}
\begin{tabular}{lllclclll}
\toprule
\textbf{Reference} & \textbf{Target Malware} & \textbf{Input Source} & \textbf{Type} & \textbf{LLM} & \textbf{Param} & \textbf{CW} & \textbf{OS} & \textbf{Accuracy} \\ \midrule

Fujii \textit{et al.}\cite{ref109} 
& Babuk ransomware 
& Decompiled/disassembled functions 
& Static 
& ChatGPT-4.0 
& - 
& 8192 
& \xmark 
& 90.90\% 
\\ \hline

\multirow{4}{*}{Simion \textit{et al.}\cite{ref116}}
& \multirow{4}{*}{General malicious files}
& \multirow{4}{*}{API call sequences}
& \multirow{4}{*}{Static}
& Llama2-13B     & 13B        & 4096 & \cmark & 50\% \\
& & & & Mistral         & 7.3B      & 8192 & \cmark & 51\% \\
& & & & Mixtral         & 7$\sim$13B & 4096 & \cmark & 67\% \\
& & & & Mixtral-FP16    & 7$\sim$13B & 4096 & \cmark & 72\% 
\\ \hline

\multirow{2}{*}{Zahan \textit{et al.}\cite{ref44}} & \multirow{2}{*}{Malicious packages} & \multirow{2}{*}{JavaScript code files} & \multirow{2}{*}{Static} & GPT-3.5-turbo-1106   & 175B & 4096 & \xmark & 91\% \\
& & & & GPT-4-1106-preview & -    & 8192 & \xmark & 99\%
\\ \hline

\multirow{3}{*}{Khan \textit{et al.}\cite{ref41}}
& \multirow{3}{*}{Android malware}
& \multirow{3}{*}{APK files}
& \multirow{3}{*}{Static}
& CodeBERT & 125M & 512  & \cmark & 95.29\% \\
& & & & GPT-2   & 1.5B & 1024 & \cmark & 94.89\% \\
& & & & RoBERTa & 125M & 512  & \cmark & 94.94\%
\\ \hline

Zhao \textit{et al.}\cite{ref42}
& Android malware
& APK files
& Static
& GPT-4-1106-preview
& -
& 8192
& \xmark
& 97.15\%
\\ \hline

\multirow{2}{*}{Yan \textit{et al.}\cite{ref119}}
& \multirow{2}{*}{General malware}
& \multirow{2}{*}{API call sequences}
& \multirow{2}{*}{Dynamic}
& BERT  & 110M & 512  & \cmark & \multirow{2}{*}{95.61\%} \\
& & & & GPT-4 & -    & 8192 & \xmark &  
\\ \hline

Sun \textit{et al.}\cite{ref120}
& Linux-based malware
& System call traces
& Dynamic
& ChatGPT-3.5
& 175B
& 4096
& \xmark
& -
\\ \hline

\multirow{6}{*}{Sanchez \textit{et al.}\cite{ref43}}
& \multirow{6}{*}{IoT malware}
& \multirow{6}{*}{System call traces}
& \multirow{6}{*}{Dynamic}
& BERT       & 110M & 512  & \cmark & 67.72\% \\
& & & & DistilBERT & 66M  & 512  & \cmark & 63\% \\
& & & & GPT-2      & 1.5B & 1024 & \cmark & 69\% \\
& & & & BigBird    & 110M & 4096 & \cmark & 87\% \\
& & & & Longformer & 150M & 4096 & \cmark & 86\% \\
& & & & Mistral    & 7.3B & 8192 & \cmark & 58\%
\\ \hline

Li \textit{et al.}\cite{ref112}
& Android malware
& \begin{tabular}[c]{@{}l@{}}Code features and system calls\end{tabular}
& Hybrid
& ChatGPT
& -
& -
& \xmark
& -
\\ \bottomrule

\end{tabular}%
\caption{Overview of the LLMs used in referenced papers, their target malware, input sources, type of analysis, parameter sizes (Param), context window sizes (CW), open-source availability (OS), and testing accuracy.}
\label{tab:malware_llm}
\end{table*}

\subsection{LLM for Malware Detection}
\label{subsec:llm_for_dmd}
As discussed in \autoref{subsec:SAMD}, the definition of malware detection is provided. This subsection focuses on using LLMs to analyze runtime data for malware detection. The distinction between static and dynamic analysis depends primarily on the input source. For instance, if API call sequences are captured during program runtime, such as through sandboxes, debuggers, or runtime analysis frameworks, they are classified as dynamic analysis. Conversely, API call sequences extracted through methods like decompilation or disassembly from static files are classified as static analysis. \autoref{tab:malware_llm} provides an overview of LLMs in both static and dynamic approaches and their testing accuracy.
\looseness=-1

\looseness=-1
Yan~\textit{et al.}~\cite{ref119} proposed a dynamic malware detection method that utilizes GPT-4 to generate text representations for API calls, which are an essential feature in dynamic malware analysis. Their method incorporates the innovative use of prompt engineering, allowing GPT-4 to generate highly detailed, context-rich descriptions for each API call in a sequence. These descriptions go beyond simple API names and delve into the specifics of how each API call behaves within the context of the malware’s execution. This provides a much deeper understanding of the malware’s actions, as opposed to traditional approaches that primarily rely on raw, unprocessed sequences of API calls. After generating these descriptions, the next step in the pipeline involves using BERT to convert the textual descriptions into embeddings. These embeddings encapsulate the semantic information of the API calls and their interactions, thereby forming a high-quality representation of the entire API sequence. These representations are then passed through a CNN, which performs feature extraction and classification. This comprehensive approach addresses several major challenges faced by traditional API-based models. 

Similarly, Sun~\textit{et al.}~\cite{ref120} developed a framework that uses dynamic analysis and LLMs to generate detailed cyber threat intelligence (CTI) reports. The framework captures syscall execution traces of malware and converts them into natural language descriptions using a Linux syscall transformer. These descriptions are organized into an Attack Scenario Graph (ASG) to preserve essential details and reduce redundancy. 
Sanchez~\textit{et al.}~\cite{ref43} applied pre-trained LLMs with transfer learning for malware detection. They fine-tuned the models with a classification layer on a dataset of benign and malicious system calls. This approach allows the model to distinguish between normal and malicious behavior while avoiding the need for training from scratch by leveraging pre-trained LLMs. 

\begin{tcolorbox}[title = {Takeaway 5},
  fonttitle = \bfseries\normalsize, 
  fontupper = \rmfamily\normalsize, 
  fontlower = \itshape]
Dynamic malware detection with LLMs analyzes runtime behaviors like API and system call traces to improve accuracy and interpretability. Larger models like GPT-4 enhance adaptability to unseen patterns, while smaller models like BERT are efficient for real-time tasks. Hybrid approaches further optimize detection by balancing interpretability and scalability.
\end{tcolorbox}

\subsection{LLM for Fuzzing}
\label{subsec:fuzzing}

\begin{figure}[t]
    \centering
    \includegraphics[width=\linewidth]{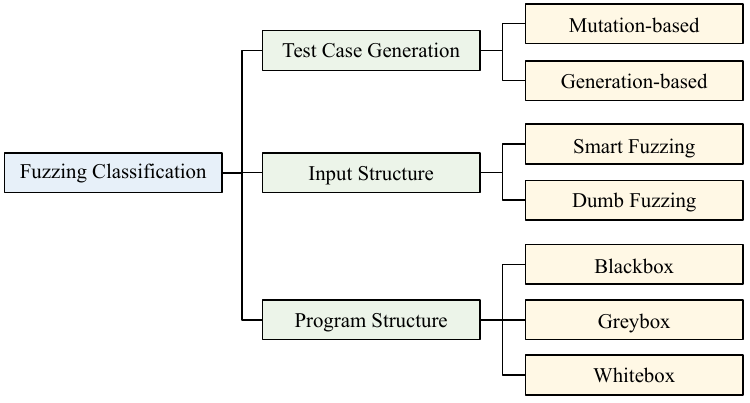} 
    \vspace{-0.2in}
    \caption{Fuzzing classifications.}
    \vspace{-0.3in}
    \label{fig:Fuzz}
\end{figure}

Fuzzing is a technique for automated software testing that inputs randomized data into a program to detect vulnerabilities like crashes, assertion failures, or undefined behaviors. The classifications of fuzzing is shown in \autoref{fig:Fuzz}. Fuzzing approaches are categorized by three dimensions: test case generation, input structure, and program structure. Test case generation can be mutation-based which alters existing inputs, or generation-based which creates new inputs from scratch. Input structure distinguishes smart fuzzing which utilizes input format knowledge, from dumb fuzzing which generates inputs blindly. Program structure analysis classifies fuzzing as black-box, grey-box, or white-box, based on the tester's level of program insight. 

The use of LLMs for fuzzing is summarized in \autoref{tab:fuzz}, which highlights the strategies, program structures, LLMs employed, and applications in the studies. Most research utilizing LLMs for fuzzing focuses on greybox fuzzing.  

\begin{table*}[]
\resizebox{\textwidth}{!}{%
\begin{tabular}{llcclcll}
\toprule
\textbf{Reference} &
  \textbf{Target} &
  \textbf{TCG} &
  \textbf{PS} &
  \textbf{LLM} &
  \textbf{Param} &
  \textbf{OS} &
  \textbf{LLMs Usage} \\ \midrule
\multirow{3}{*}{CHEMFuzz\cite{ref70}} &
  \multirow{3}{*}{\begin{tabular}[c]{@{}l@{}}Quantum chemistry \\ software\end{tabular}} &
  \multirow{3}{*}{Mutation} &
  \multirow{3}{*}{Greybox} &
  GPT-3.5 &
  175B &
  \xmark &
  \multirow{3}{*}{\begin{tabular}[c]{@{}l@{}}Input file mutation and output \\ analysis\end{tabular}} \\
 &
   &
   &
   &
  Claude-2*\cite{ref133} &
  - &
  \xmark &
   \\
 &
   &
   &
   &
  Bard\cite{ref134} &
  - &
  \xmark &
   \\ \hline
CovRL\cite{ref71} &
  JavaScript Engines &
  Mutation &
  Greybox &
  CodeT5+ &
  220M &
  \cmark &
  Generates valid test cases \\ \hline
LLAMAFUZZ\cite{ref73} &
  Real-world programs &
  Mutation &
  Greybox &
  llama-2-7b-chat-hf &
  7B &
  \cmark &
  \begin{tabular}[c]{@{}l@{}}Mutate structured data inputs \\ and generate new seeds\end{tabular} \\ \hline
\multirow{2}{*}{FuzzGPT\cite{ref138}} &
  \multirow{2}{*}{Deep Learning Libraries} &
  \multirow{2}{*}{Mutation} &
  \multirow{2}{*}{Greybox} &
  Codex (code-davinci-002) &
  - &
  \xmark &
  Mutatie and refine test cases \\
 &
   &
   &
   &
  CodeGen (350M/2B/6B-mono) &
  350M/2B/6B &
  \cmark &
  Generates initial test cases \\ \hline
CHATFUZZ\cite{ref141} &
  General programs &
  Mutation &
  Greybox &
  GPT-3.5-turbo &
  175B &
  \xmark &
  \begin{tabular}[c]{@{}l@{}}Generates format-conforming \\ variations of existing seeds\end{tabular} \\ \hline
CODAMOSA\cite{ref142} &
  Python modules &
  Mutation &
  Greybox &
  Codex &
  - &
  \xmark &
  \begin{tabular}[c]{@{}l@{}}Generates tailored inputs and\\  extends callable sets\end{tabular} \\ \hline
CHATAFL\cite{ref143} &
  \begin{tabular}[c]{@{}l@{}}Network protocol \\ implementations\end{tabular} &
  Mutation &
  Greybox &
  GPT-3.5-turbo &
  175B &
  \xmark &
  \begin{tabular}[c]{@{}l@{}}Extracts grammars and enrich\\  seed corpora\end{tabular} \\ \hline
Asmita\textit{et al.}\cite{ref144} &
  BusyBox &
  Mutation &
  \begin{tabular}[c]{@{}c@{}}Greybox,\\ blackbox\end{tabular} &
  GPT-4-0613 &
  - &
  \xmark &
  Generate seeds \\ \hline
Fuzz4All\cite{ref72} &
  \begin{tabular}[c]{@{}l@{}}Compilers, SMT solvers, \\ quantum frameworks and \\ programming toolchains.\end{tabular} &
  \begin{tabular}[c]{@{}c@{}}Mutation,\\  generation\end{tabular} &
  Greybox &
  GPT-4.0 &
  - &
  \xmark &
  Generates  fuzzing inputs \\ \bottomrule
\end{tabular}%
}
\caption{Overview of the LLM-based fuzzers used in referenced papers, including their target software, test case generation (TCG), program structure (PS), model parameters, open-source availability (OS), and usage details.}
\label{tab:fuzz}
\end{table*}

Qiu~\textit{et al.}\cite{ref70} introduced CHEMFuzz, an LLM-assisted fuzzing framework designed for quantum chemistry software. CHEMFuzz uses an evolutionary fuzzing approach with LLM-based input mutation and output analysis to address the syntactic and semantic complexities of quantum chemistry software. The two-module system combining syntactic mutation operators with anomaly detection detected 40 bugs and 81 potential warnings in Siesta 4.1.5\cite{ref135}. 
Eom~\textit{et al.}~\cite{ref71} introduced CovRL, a  framework that integrates coverage-guided reinforcement learning with LLMs to enhance fuzzing for JavaScript engines. The approach combines Term Frequency-Inverse Document Frequency (TF-IDF) weighted coverage maps with reinforcement learning to guide the LLM-based mutator. This enables the generation of more effective test cases, discovering new coverage areas and improving the efficiency of JavaScript engine fuzzing.
Deng~\textit{et al.}~\cite{ref138} introduced FuzzGPT, a framework for fuzzing deep learning libraries. By mining historical bug-triggering programs and leveraging LLMs such as Codex\cite{ref139} and CodeGen~\cite{ref140}, FuzzGPT generates edge-case inputs using strategies like few-shot, zero-shot, and fine-tuned learning. This targeted approach exploits API-specific vulnerabilities, illustrating the effectiveness of LLMs in managing complex software ecosystems.
Meng~\textit{et al.}~\cite{ref143} introduced CHATAFL, an LLM-guided mutation-based framework for protocol fuzzing. The framework extracts protocol grammars, enhances seed diversity, and transitions to unexplored protocol states. This approach overcomes challenges like reliance on initial seeds and restricted state-space exploration.
\looseness=-1

Beyond domain-specific applications, frameworks like LLAMAFUZZ~\cite{ref73} and CHATFUZZ~\cite{ref141} showcase the adaptability of LLMs for general program fuzzing. Zhang~\textit{et al.} proposed LLAMAFUZZ, which combines greybox fuzzing with LLM-based mutation to enhance branch coverage and bug detection. Its focus on structured data inputs makes it an effective tool for augmenting traditional fuzzing methods, demonstrating improvements over AFL++~\cite{ref74}. Similarly, Hu~\textit{et al.} introduced CHATFUZZ, leveraging ChatGPT to generate format-conforming test cases for highly structured inputs, addressing the efficiency limitations of traditional mutation- and grammar-based fuzzers. These frameworks demonstrate the ability of LLMs to adapt to structured program requirements while advancing fuzzing efficiency.

Lemieux~\textit{et al.}~\cite{ref142} introduced CODAMOSA, an approach that integrates LLMs into testing workflows. CODAMOSA combines Search-Based Software Testing (SBST) with Codex~\cite{ref139} to generate test cases and address coverage stagnation. It integrates LLM-generated Python code into SBST workflows, highlighting the collaboration between traditional testing and LLM-driven techniques. Asmita~\textit{et al.}~\cite{ref144} explored LLM-based fuzzing in BusyBox~\cite{ref145}, a widely used Linux utility suite. Their approach combines LLM-assisted seed generation with crash reuse to enhance efficiency in black-box fuzzing workflows. Using GPT-4, they demonstrated how LLMs handle complex inputs and reuse crashes for cross-variant testing, improving vulnerability detection. Additionally, Xia~\textit{et al.}~\cite{ref72} proposed Fuzz4All, a universal fuzzing framework that extends fuzzing beyond language- or system-specific constraints. Fuzz4All uses autoprompting and an iterative fuzzing loop to transform user-provided inputs into prompts for generating diverse test cases.

\begin{tcolorbox}[title = {Takeaway 6},
  fonttitle = \bfseries\normalsize, 
  fontupper = \rmfamily\normalsize, 
  fontlower = \itshape,
  breakable]
LLM-based fuzzing frameworks have advanced automated testing by combining mutation-based and generation-based strategies with models like GPT-3.5, Codex, and CodeGen. As shown in \autoref{tab:fuzz}, these tools share common goals, such as improving test coverage, addressing domain-specific challenges, and automating seed generation and refinement.  
\end{tcolorbox}

\subsection{LLM for Penetration Testing}
\label{subsec:llm_for_pt}
Penetration testing is a controlled security assessment that simulates real-world attacks to identify, evaluate, and mitigate vulnerabilities in systems and networks\cite{ref76}.

Deng~\textit{et al.}~\cite{ref77} explored the capabilities of LLMs in penetration testing, revealing that while these models excel at sub-tasks, they face challenges in maintaining context across multi-step workflows. To address this limitation, the authors proposed PentestGPT, a framework integrating reasoning, generation, and parsing modules. This framework significantly improved task completion rates by 228.6\% compared to GPT-3.5 and demonstrated effective performance in real-world scenarios.
Huang~\textit{et al.}~\cite{ref78} developed PenHeal, an LLM-based framework combining penetration testing and remediation. PenHeal includes a Pentest Module that uses techniques like counterfactual prompting to autonomously detect vulnerabilities. Its remediation module offers tailored strategies based on severity and cost efficiency. Compared to PentestGPT, PenHeal increased detection coverage by 31\%, improved remediation effectiveness by 32\%, and reduced costs by 46\%.
Additionally, Goyal~\textit{et al.}~\cite{ref147} proposed Pentest Copilot, a framework that uses GPT-4-turbo to enhance penetration testing workflows. Pentest Copilot incorporates chain-of-thought reasoning and retrieval-augmented generation to automate tool orchestration and exploit exploration. It ensures adaptability with a web-based interface. This approach combines automation with expert oversight, enhancing the accessibility of penetration testing while preserving technical depth.

\begin{figure*}[t] 
    \centering 
    \includegraphics[width=0.99\textwidth]{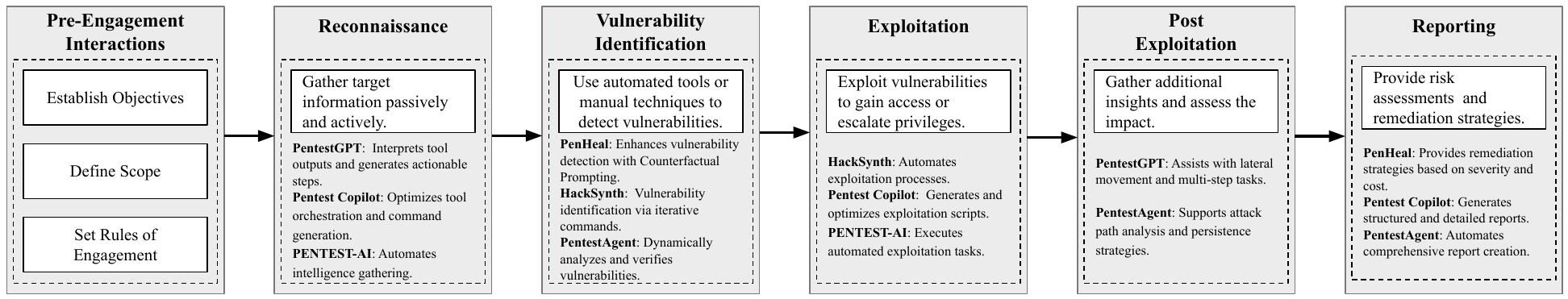} 
    \caption{Integration of LLMs across the six steps of penetration testing.}
    \vspace{-0.25in}
    \label{fig:Pentest} 
\end{figure*}

\looseness=-1
Additionally, some frameworks are designed as agent-based systems. Bianou~\textit{et al.}~\cite{ref148} presented PENTEST-AI, a framework guided by the MITRE ATT\&CK framework for multi-agent penetration testing. The framework automates reconnaissance, exploitation, and reporting tasks using specialized LLM agents. PENTEST-AI reduces human intervention while aligning with established cybersecurity methodologies, illustrating the synergy between LLMs and structured security frameworks in addressing real-world challenges.
Muzsai~\textit{et al.}~\cite{ref149} proposed HackSynth, an LLM-driven penetration testing agent with two modules: a Planner for generating commands and a Summarizer for processing feedback. Tested on newly developed CTF-based benchmarks, HackSynth demonstrated its capability to autonomously exploit vulnerabilities and achieve optimal performance with GPT-4. 
Gioacchini~\textit{et al.}\cite{ref146} developed AutoPenBench, a framework with 33 tasks covering experimental and real-world penetration testing scenarios. AutoPenBench compares autonomous and semi-autonomous agents, tackling reproducibility challenges in penetration testing research. Fully autonomous agents achieved a 21\% success rate, significantly lower than the 64\% success rate of semi-autonomous setups. 
Shen~\textit{et al.}~\cite{ref150} introduced PentestAgent, leveraging LLMs and Retrieval-Augmented Generation (RAG) to automate intelligence gathering, vulnerability analysis, and exploitation. PentestAgent dynamically integrates tools and adapts to diverse environments, improving task completion and operational efficiency. It outperforms existing LLM-based penetration testing systems.

As illustrated in \autoref{fig:Pentest}, penetration testing involves six stages: pre-engagement interactions, reconnaissance, vulnerability identification, exploitation, post-exploitation, and reporting. Pre-engagement interactions establish objectives, define scope, and set rules of engagement. Reconnaissance gathers target information through passive and active methods to identify attack vectors. Vulnerability identification uses automated tools and manual techniques to detect and verify weaknesses. Exploitation leverages these vulnerabilities to demonstrate potential risks, while post-exploitation assesses the breach's impact and ensures persistence if needed. Finally, reporting consolidates findings into structured documentation with risk assessments and remediation strategies.

\begin{tcolorbox}[title = {Takeaway 7},
  fonttitle = \bfseries\normalsize, 
  fontupper = \rmfamily\normalsize, 
  fontlower = \itshape,
  breakable]
LLMs can be applied across multiple stages of penetration testing. For example, LLM-driven frameworks simplify reconnaissance by automating tool output interpretation and intelligence gathering. They improve vulnerability identification through dynamic analysis methods, including counterfactual prompting. Additionally, LLMs assist in post-exploitation by facilitating multi-step attack strategies.
\end{tcolorbox}

\section{LLM for Hybrid Approach}
\label{sec:ha}
A hybrid approach employs both static and dynamic analysis techniques at different stages. For example, combining static features like code structure or permissions with dynamic behaviors such as system calls or memory usage represents a hybrid approach. This section discusses the role of LLMs in hybrid approaches, focusing on two aspects: (i) LLM for unit test generation (\autoref{subsec:llm_for_utg}) and (ii) other hybrid methods (\autoref{subsec:hybrid_others}). 

\subsection{LLM for Unit Test Generation}
\label{subsec:llm_for_utg}
Unit testing is a fundamental practice in software development that focuses on verifying the functionality of individual components or "units" of a program. By isolating and testing each unit, developers can ensure code correctness, detect errors early, and improve overall code quality. Traditional unit test generation methods are written manually by developers and generally involve search-based, constraint-based, or random techniques to maximize code coverage ~\cite{ref47}. Automated unit test generation leverages tools and techniques to generate tests automatically, reducing developer workload and improving coverage. Static analysis is essential in guiding test generation by examining the program's structure, dependencies, and control flow. Dynamic analysis complements this by evaluating the generated tests through runtime execution, identifying errors, and refining test quality. Together, these hybrid approaches enhance the efficiency and effectiveness of unit test generation.

\textbf{Performance Comparison Between LLMs and Traditional Test Generation Tools.} A study evaluated the performance of ChatGPT and Pynguin~\cite{ref151} in generating unit tests for Python programs, focusing on three types of code structures: procedural scripts, function-based modular code, and class-based modular code. Bhatia~\textit{et al.}~\cite{ref152} compared the tools in terms of coverage, correctness, and iterative improvement through prompt engineering. They found that ChatGPT and Pynguin achieved comparable statement and branch coverage. Iterative prompting improved ChatGPT’s coverage for function- and class-based code, saturating after four iterations, but showed no improvement for procedural scripts. The study also revealed minimal overlap in missed statements, suggesting combining the tools could enhance coverage. However, ChatGPT often generated incorrect assertions, especially for less structured code, due to its focus on natural language over code semantics. The authors concluded that while LLMs like ChatGPT are promising for unit test generation, integrating semantic understanding and combining them with traditional tools could address current limitations and improve performance.

\begin{figure}[t]
    \centering
    \includegraphics[width=\linewidth]{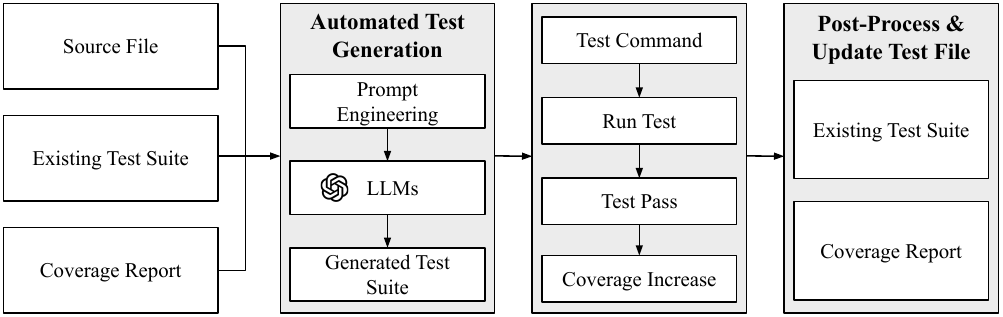} 
    \caption{Workflow of unit test generation with LLMs.}
    \vspace{-0.25in}
    \label{fig:Unit}
\end{figure}

\textbf{Static Analysis-Assited Unit Test Generation.} One improvement is the ability of LLMs to generate focused and meaningful test cases by using static analysis to extract and structure relevant context. For instance, aster~\cite{ref153} and ChatUniTest~\cite{ref46}  integrate techniques such as dependency extraction, program slicing, and adaptive focal context. These methods ensure that prompts sent to LLMs are concise and focused, enabling the generation of tests that better align with the target methods. Similarly, APT~\cite{ref155} employs a property-based approach to guide LLMs in generating tests using the "Given-When-Then" paradigm, which improves logical structure in generated tests. These static analysis techniques address the limitations of traditional methods, which often struggle to extract relevant dependencies or fail to focus on critical components of the code.

\textbf{Dynamic Analysis-Assited Unit Test Generation.} Dynamic analysis complements static techniques by validating and refining test cases through iterative processes, improving coverage and correctness. For example, TestART~\cite{ref154} uses a co-evolutionary framework to iteratively generate and repair tests based on runtime feedback, addressing flaky or invalid tests often produced by traditional methods. In ChatUniTest~\cite{ref46}, dynamic validation integrates runtime error detection with rule-based and LLM-driven repair, ensuring that generated tests are compilable and logically sound. Furthermore, ChatTester~\cite{ref157} demonstrates how iterative prompting based on dynamic feedback can address missed statements and branches, progressively improving line and branch coverage. These dynamic techniques allow LLM-based approaches to adapt and refine tests, addressing limitations of traditional static tools that lack iterative capabilities.

\textbf{Prompt Engineering.} Techniques like adaptive focal context in ChatUniTest~\cite{ref46} and program slicing in HITS~\cite{ref158} streamline prompts by reducing irrelevant information, ensuring the LLM remain focused. Chain-of-thought reasoning, as seen in aster~\cite{ref153}, enhanced the LLM’s ability to handle complex dependencies and generated logically coherent tests. Additionally, AGONETEST~\cite{ref159} employed structured prompts incorporating mock dependencies and example inputs, guiding the LLM to generate more comprehensive test cases. These techniques address the inflexibility of traditional tools, which often rely on predefined templates and lack the ability to dynamically adapt prompts based on code context.

\begin{tcolorbox}[title = {Takeaway 8},
  fonttitle = \bfseries\normalsize, 
  fontupper = \rmfamily\normalsize, 
  fontlower = \itshape,
  breakable]
Static and dynamic analysis operate at distinct stages in unit test generation. Static analysis extracts dependencies and slices programs, enabling LLMs to generate targeted, logically structured tests. Dynamic analysis then validates and refines these tests through runtime feedback. Prompt engineering techniques such as adaptive focal context and structured prompts, align test generation with code semantics to enhance coverage.
\end{tcolorbox}

\subsection{Others}
\label{subsec:hybrid_others}
In addition to the previously discussed methods for unit test generation, other hybrid approaches integrate static and dynamic analysis through an agent framework. This framework first performs static analysis, such as extracting ASTs and analyzing code structure, and then conducts dynamic testing.

\textbf{Multi-Agent Framework for Secure Code Generation.} Nunez~\textit{et al.}~\cite{ref81} introduced AutoSafeCoder, an innovative multi-agent framework designed to improve the security of automatically generated code. The framework leverages three distinct LLM-driven agents working collaboratively to generate, analyze, and secure code. The Coding Agent is responsible for generating the initial code, while the Static Analyzer Agent identifies potential vulnerabilities through AST analysis. Meanwhile, the Fuzzing Agent detects runtime errors by employing mutation-based fuzzing techniques, ensuring that the generated code performs securely during execution. Interactive feedback loops integrate both static and dynamic testing methods into the code generation process, optimizing the outputs from the LLM at each stage. 

\textbf{Coverage Test Generation.} Pizzorno~\textit{et al.}~\cite{ref82} presented CoverUp, a method for generating Python regression tests with high code coverage. CoverUp evaluates existing code coverage, identifies gaps, and uses LLMs to generate new tests informed by static analysis. If tests fail to execute or enhance coverage, CoverUp iteratively refines them using error messages and code context. This process continues until all segments are fully tested and integration issues are resolved.

\textbf{Malware Analysis.}  Li~\textit{et al.}~\cite{ref112} used reverse engineering tools to extract static and dynamic features from Android APK files, organizing them into permissions, system calls, and metadata. They used tailored prompts to guide ChatGPT in generating textual analyses and maliciousness scores for each application. These results were compared with three existing Android malware detection models: Drebin~\cite{ref113}, MaMaDroid~\cite{ref114}, and XMAL~\cite{ref115}. Although traditional models showed strong classification capabilities, the authors noted their limitations in interpretability and dataset dependency. ChatGPT offered comprehensive analyses and explanations but lacked decision-making capabilities.

\textbf{Malware Reverse Engineering.} Williamson~\textit{et al.}~\cite{ref83} integrated LLMs with static and dynamic analysis techniques to enhance malware reverse engineering. In the static phase, tools like IDA Pro examined binaries to extract structural details such as embedded strings and control flow. In the dynamic phase, sandboxes monitored malware behavior, capturing network and system interactions. LLMs synthesized results from both phases, deriving actionable insights and identifying indicators of compromise (IoCs). 

\begin{tcolorbox}[title = {Takeaway 9},
  fonttitle = \bfseries\normalsize, 
  fontupper = \rmfamily\normalsize, 
  fontlower = \itshape,
  breakable]
LLMs enhance hybrid methods by iteratively refining the outputs of static (e.g., AST analysis in AutoSafeCoder, coverage gaps in CoverUp) and dynamic (e.g., fuzzing, runtime feedback) analysis process. They bridge code structures with runtime behaviors, enabling secure code generation, high-coverage tests, and actionable malware analysis.
\end{tcolorbox}

\section{Discussion}
\label{sec:discussion}
The use of LLMs in the field of program analysis has  mitigated several previous limitations such as false positives, performance overhead, inherent knowledge barriers, path explosion, the trade-off between speed and accuracy, and the difficulty of achieving automation across diverse systems without heavy manual intervention. Despite these advancements, new limitations and challenges have emerged with the introduction of LLMs. The following subsections provide an overview of these challenges (\autoref{subsec:cha}) and discuss potential future research directions (\autoref{subsec:future}).

\subsection{Challenges}
\label{subsec:cha}
\textbf{Technical Limitations.}
LLMs face several technical challenges in program analysis. First, incorrect data type identification and information loss during decompilation reduce analysis accuracy. Second, LLMs often oversimplify patches, limiting their ability to address vulnerabilities in real-world applications. In some cases, they produce empty responses, particularly during software verification and patching tasks. Third, LLMs struggle with variable reuse, often confusing identically named variables in different scopes. Finally, LLMs struggle to analyze logic vulnerabilities involving intricate control flows, complex nesting, and time-based competition conditions. These challenges reduce their effectiveness in assessing such scenarios.

\textbf{Model Characteristics and Limitations.}LLMs are non-deterministic and may produce varying outputs for identical inputs, complicating consistency in repeated vulnerability assessments. This variability hinders reliable and repeatable results. Additionally, LLMs are prone to hallucinations, generating fabricated information that misleads vulnerability detection. These limitations in consistency and accuracy make LLMs insufficient for reliable program analysis.
\looseness=-1

\textbf{Cost and Dependency Issues.}
The effectiveness of LLM-based program analysis relies on prompt engineering, which requires significant expertise. Poorly designed prompts can lead to ineffective results or introduce biases, limiting the model's ability to detect vulnerabilities. Furthermore, using LLMs can be costly, especially when analyzing long code segments, due to the large number of tokens required. The inherent token limits of LLMs also restrict their ability to handle extensive or complex programs, making scalability a challenge in real-world applications.

\subsection{Future Directions}
\label{subsec:future}
\textbf{Deep Integration of LLMs with Analysis Techniques. }
Most current methods use LLMs independently of program analysis. Integrating LLMs with static analysis into a unified workflow offers opportunities for enhanced effectiveness. Some studies~\cite{ref33} have acknowledged that their methods lack effective integration of LLMs with other models or techniques. Frameworks combining LLMs with GNNs~\cite{ref31} for program control and data flow have shown significant improvements in detection accuracy. Future work should focus on integrating LLMs with static and dynamic analysis to create more effective solutions for vulnerability detection.

\textbf{Transforming Dynamic Analysis into Static Analysis.} Transforming tasks traditionally requiring dynamic analysis into static analysis with LLMs is an emerging direction. Tasks like runtime vulnerability detection and memory corruption analysis historically depended on dynamic analysis to capture execution-specific behaviors. LLM integration can shift these processes to static analysis, enabling early vulnerability detection without runtime execution. This reduces computational overhead, avoids repeated executions, and improves scalability for analyzing large systems. Pei~\textit{et al.}~\cite{ref84} showed how fine-tuning LLMs eliminates the need for runtime information by predicting program invariants from source code, enabling earlier safety checks during compilation.
\looseness=-1

\textbf{Emulating Human Security Researchers for Vulnerability Detection.} Advancing code understanding and reasoning capabilities enable LLMs to replicate systematic approaches used by human security researchers. LLMs overcome the rule-based limitations of traditional tools by analyzing complex code contexts and identifying nuanced vulnerabilities. This enables LLMs to mimic hypothesis-driven processes, identifying subtle vulnerabilities missed by automated methods. Glazunov~\textit{et al.}~\cite{ref85} introduced Project Naptime to replicate human security researchers' workflows for vulnerability detection. The framework employs tools such as a code browser, Python interpreter, and debugger, enabling LLMs to perform expert-level code analysis and vulnerability detection. Evaluated on the CyberSecEval 2\cite{ref86} benchmark, this approach improves detection and demonstrates the feasibility of automating complex security tasks.

\section{Conclusion}
\label{sec:conclusion}
Integrating LLMs into program analysis enhances vulnerability detection, code comprehension, and security assessments. LLMs’ natural language processing capabilities, combined with static and dynamic analysis techniques have improved automation, scalability, and interpretability in program analysis. These advancements facilitate faster vulnerability detection and provide deeper insights into software behavior. Challenges such as token limitations, path explosion, complex logic vulnerabilities, and LLM hallucinations remain barriers. The studies reviewed in this survey highlight recent progress, offering insights into its current state and emerging opportunities. Future directions include developing domain-specific models, refining hybrid methods, and enhancing reliability and interpretability to fully utilize LLMs in program analysis. This survey aims to assist in addressing the mentioned challenges and inspire the development of more effective program analysis frameworks.
\looseness=-1

\bibliographystyle{IEEEtran}
\bibliography{reference}

\end{document}